# *Engaged pedagogy: An Innovative method to Teach Physics*

*Iulia Salaoru*

*Emerging Technologies Research Centre*

*De Montfort University,*

*The Gateway, Leicester, LE1 9BH, United Kingdom*

*"Education is the kindling of a flame, not the filling of a vessel."*

*Socrates*

**Abstract**

Most science classes and in particular Physics are delivered through traditional teaching methods. More specifically, in the traditional stand-and-deliver lecture, the information was transmitted unilaterally from the teacher to the student and hence the students are only passive participants. Recent studies revealed that traditional lectures are inefficient and even more do not help students understand key concepts or develop critical thinking skills. Mazur [1] described the traditional lecture as "*the illusion of teaching for teachers, and the illusion of learning for learners*". However, one of the approaches that can support students to study Physics more efficiently as well as develop critical and quantitative thinking skills is *Engaged pedagogy*. Engaged pedagogy, as a novel interactive teaching method and as well as a vector of success in teaching and learning Physics, is discussed in this paper. Additionally, the impact of technology on the proposed teaching pathway is presented.

## 1. Introduction

Most learners make their first contact with Physics as a fundamental science that helps them understand how everything around works, from the microcosm of cars, light bulbs, computers, mobile phones or DNA to the macrocosm, for instance the Bing Bang theory of universe formation [2]. At this stage, **Interesting**, **Useful** and **Challenging** are three powerful words that could describe Physics as a science. **Interesting** – as Physics has the potential to interest and enthuse learners of all ages due to its strong relevance to everyday life. **Useful** – as Physics provides the explanations of the concepts/phenomena behind the tools/equipment and instruments used by engineers, doctors and other professionals [3]. Yet, Physics is **challenging** as, especially as learners progress towards more advanced levels, significant effort is required to understand its fundamental concepts as well as to develop problem solving skills. As a result, Physics comes to be labelled as a hard-to-learn subject. Under these circumstances, relatively small numbers of students choose to study Physics at degree level; it should be emphasized that the teacher and teaching method(s) are crucial in maintaining these students' interest, in providing accessible and engaging classes and at the same time supporting the development of their learning [4,5].

Recent studies revealed that the traditional stand-and-deliver lectures with limited student participation are not only boring and difficult to follow but also inefficient, as they do not help students develop their understanding of concepts or fundamental processes [6-9]. More specifically, in a traditional lecture, information is transmitted unilaterally from the teacher to the receiver/student as is illustrated in Figure 1. Additionally, in this lecture style, the students have only a passive role and the teachers are bombarding the students with information without any feedback from them. Damodharan et al. [10] revealed that during a traditional lecture, students' concentration fades after 15-20 minutes.

In order to help students study Physics more efficiently as well as to develop critical and quantitative thinking skills, an alternative method to the traditional one was proposed, called **Active learning**. Notably, this alternative approach helps students study Physics in interesting and enjoyable ways. In this teaching style, students are converted from passive listeners into active participants to the lecture (Figure 1). Indeed, Freeman [11] and his collaborators conducted a study on 225 students to investigate the impact of lecture type (traditional vs. active learning) on learning and course performance. This study demonstrated that the active learning lecture could be linked to a reduction in the failure rates by 1.5 times when compared to the traditional approach and as well as an improvement in average examination grades by about 6%.

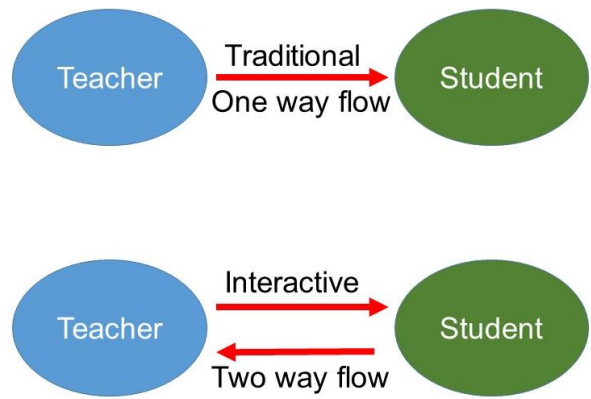

Fig.1 Teacher-student interactions

Currently, one of the strategic goals in Higher Education (HE) institutions is to improve the quality and efficiency of the students' teaching and learning experience. In light of this, we can highlight that in general and with regards to the discipline of Physics in particular, the most effective teaching approach is the interactive-engagement pedagogy [12,13]. As a result of this, the students could be more effectively engaged in the learning process and hence their overall performance, learning and skills could be maximized.

The aim of this paper is to discuss the potential of engaged pedagogy as a vector of success in teaching and learning Physics. The Learning Technology tools involved in the proposed teaching pathway will also be addressed.

## 2. Engaged pedagogy

The term 'engaged pedagogy' was introduced in 1994 by Bell Hooks, an African American social activist. Engaged pedagogy is a teaching method which postulates that lectures/classes should never be boring, but a source of interest and excitement, with learners actively involved in the learning process. In fact, Bell Hooks emphasizes that it is important to deliver information, but it is more important how you do that: *"Our work is not merely to share information but to share in the intellectual and spiritual growth of our students"*[14,15]. In other words, engaged pedagogy is an interactive two-way flow of information where both teachers and students are involved passionately with academic material (see Figure 1). Engaged pedagogy aims to put an end to apathy, disinterest and boredom during classes and empower students to assume responsibility for their learning. The sections below provide an overview of different methods that could be used to implement engaged pedagogy in the Physics classroom.

*2.1 First day of class* The first lecture is crucial as it sets the tone for the subsequent sessions. During the first lecture, the teacher has the opportunity to share their interest, enthusiasm and excitement about the subject, incite students' curiosity regarding the topics that will be covered and also create a positive and relaxed classroom atmosphere. Notably, the teacher has a crucial role at this point and if he/she has a confident attitude, is enthusiastic and easy to follow, then the students' potential anxiety and uncertainty will be appeased [16,17]. In this respect, involvement of students in discussions about module content/topics is a very important aspect. Another essential strategy is to focus on boosting students' learning motivation. For example, a Physics lecturer could attempt to make topics more attractive to students by placing them in a context they can relate to, for instance by discussing their impact on different fields i.e. medicine, engineering or even everyday life.

*2.2 Using a historical approach.* This method helps the students find out when a Phenomenon/hypothesis/concept was discovered and then understand its evolution over time. Both March [18] and Duhem [19] promoted the historical method in teaching Physics: "*The legitimate, sure and fruitful method of preparing a student to receive a physical hypothesis is the historical method*". This approach helps students understand that the discovery of a phenomenon is important, but its evolution and further improvement are equally important. Additionally, they can see an example of scientific progress and also understand that they can have an active role in further development. Furthermore, the historical approach may be used to teach Physics in relation to the philosophical ideas that led to the development of scientific theories [20]. In this respect, an open discussion can be created around a physical concept, for example light or relativity, and the students can express their understanding and clarify their misunderstandings around this concept. Using this teaching approach, students can also be introduced to the pathway followed by an individual scientist.

In a parallel vein, it is important to highlight the universality of Physics as a discipline for all people irrespective of their gender, country or race. Actually, the historical approach provides a pathway to introduce students to Physics as a culture. Ideally here the teaching environment can be created using media technology such as films, documentaries and videos from You-Tube or scientific TV i.e. Discovery Channel.

2.3. *Interactive lectures.* This method provides activities that can give the opportunity for students to have an active role during classes. This method is focussed on learning/helping students rather than on simple delivery of material. Moreover, the students switch from passive notes-takers into active, de facto-teachers. In fact the lecture should be a mode to transfer knowledge/information from the teacher to the students' mind not only to their notebooks. In reference to conventional teaching, Eric Mazur, Professor of Physics and Applied Physics at Harvard University, states: "*Sitting passively and taking notes is just not a way of learning*" [1]. Similarly, philosopher Albert Camus reflects that: "*Some people talk in their sleep. Lecturers talk while* other *people sleep*" [21]. The impact of the interactive lecture was presented in a very powerful way in Harvard Magazine by Craig Lambert [1], who welcomes the fact that the 600-year style of traditional lecturing is finally being replaced by "interactive teaching". Lambert refers to Mazur's claim that no matter how good a teacher might be, students do not fully understand fundamental concepts through conventional lectures. Mazur performed an experiment by asking his students to have a three-minute discussion on one of the concepts he had taught and he was amazed that in this short interval the students figured out the correct answer. Additionally, he found out that the delivery way and the student learning pathway do not always match. In support of this statement, we can consider the question that Mazur received from one of his students: *"How should I answer these questions—according to what* you *taught me, or how I usually think about these things? [1]"*. On the other hand, interactive teaching is a two-way flow of information between teacher and students, and that is why this teaching pathway is a very powerful one. Another important finding derived from Mazur's exercise refers to the long-term retention of knowledge gained in interactive lectures. In contrast, "*In a traditional Physics course, two months after taking the final exam, people are back to where they were before taking the course*" [1].

We can conclude that it is imperative for teachers to design and deliver interactive lectures to support students' learning. Indeed, there are a large number of techniques/tools such as: interactive lecture demonstrations, just-in-time teaching (JiTT), using classroom engagement software, Socratic questioning and cooperative learning.

2.3.1. *Interactive lecture demonstrations* is a teaching method where the students are interactively engaged in lectures. This method is highly suitable to teaching Physics, as Physics is a practical subject. For example, the teacher prepares an experiment that is related to the concepts taught during the lecture. The students are initially invited to predict its outcome, then, as the teacher performs the experiment, they observe the result and finally discuss the physical concept behind the experiment; initially individually and then with a partner or in a group. Using a live demonstration is an effective path to illustrate a new subject/problem and develop students' ability to observe an experiment, collect data, derive physical theories, make a logical connection between them and be confident to discuss their understanding with colleagues [22]. In addition to real-time experiment(s), a range of technologies (computer-assisted data acquisition device; simulation software of different physical processes) can be used to support this teaching pathway. Based on the evidence presented above, this interactive method can support students to retain the experiment and its outcome(s) for a long time.

From a subject-specific perspective, this method is very effective as students can achieve a better understanding of physical laws, theories and concepts, connect theory with practical experiments and develop their interest in learning Physics. Topic such as gravity, heat and temperature, simple electric circuits, lenses and image formation and many more can be explored by means of this teaching tool.

2.3.2. The *Just-in-time teaching (JiTT)* method was developed by Gregor Novak and Andy Gavrin at Indiana University/Purdue University [23]. This method brings together the lecture, group discussions/problem solving and technology. Thus, the JiTT method aims to enhance student learning by asking students to read a material and to respond to a set of questions in advance of the lecture. In this way students are encouraged to learn independently, which can lead to improved engagement. Subsequently, the teacher reviews students' responses and hence becomes aware of the level of the students' understanding of the material before the lecture. This allows them to design activities that address the learning gaps identified in students' responses [23]. This teaching method can be very effective in improving students' learning as students are directly engaged in the teaching process. Interestingly, this method was originally developed in 1999 to be used in support of teaching Physics. Electronic technology is employed in this teaching method as a tool for communication [24]. More specifically, the recommended materials/questions (book chapter; video, simulation or data) are delivered via an electronic communications network. The same electronic platform is used to collect responses from students and then to provide feedback. This quick exchange of information is one of the main advantages of using electronic technology.

2.3.3 *Classroom engagement technology* In this method, an electronic device is employed to capture the students' processes of thinking by prompting them to answer a series of questions during the lecture [25]. One of the devices that can be used is the iClicker. As in Just-in-time teaching method, the question protocol is used to facilitate student engagement. In contrast to JiTT, in this case, the questions are answered during the lecture time, thus supporting the interactive aspect of the delivery [26]. Depending upon student response, brief questions can be followed up with more in-depth discussions, addressing any gaps in understanding. As previously stated, Mazur [1] observed that students' problem solving abilities significantly increased after only 2 minutes of discussion. Notably, this teaching tool can be used with large student groups and can be particularly effective in engaging students who are otherwise anxious of contributing to whole class discussions. However, suitable technology (e.g. relevant software and hand-held devices) needs to be available for the implementation of this teaching method. The software should provide a platform that not only enables the provision of answers to questions but also the timely display and analysis of responses.

2.3.4 *Socratic Questioning* This active engagement teaching method was derived from a pedagogical approach developed by the famous Greek philosopher Socrates. As previously discussed, in the traditional lecture, the teacher delivers the material and the students are passively taking notes. The Socratic questioning approach [27] is a powerful teaching tool to switch students from passive viewers to active participants in the teaching process and in the same time to help them develop critical thinking and problem solving skills. In lectures designed using this approach, teacher-centred delivery is replaced with a student-led discussion based on a set of questions related to a specific topic. Students are involved directly in asking and answering questions and hence this pathway helps develop deep learning and critical thinking skills. A quote from Socrates can help frame the reasons beyond the effectiveness of this method: "*To find yourself, think for yourself*" [28]. A number of electronic resources such as websites, video and audio material can support this teaching method.

From a subject-specific perspective, most topics included in the Physics university curriculum, from Mechanics (e.g. force, friction) to Quantum Physics (e.g. Schrödinger equation) can be taught using this method.

*2.3.5. Cooperative learning* is an educational approach where the students work in small groups to achieve a common objective (e.g. solve complex problems that are very difficult to be solved individually) [13]. Yet, it should be highlighted that every member of the group contributes to the group's success and vice-versa. Indeed, this approach aims to facilitate the development of learning for all members of the group rather than isolated individuals. Moreover, it is expected that members of the group work effectively together through discussions, resource sharing and mutual support. Cooperative learning can take a variety of pathways: in or out of class discussion or study groups, project groups and lab groups. Cooperative learning can also be implemented within lectures. This pathway is most suitable when a difficult problem has been taught and further clarification is necessary. In teaching empirical sciences such as Physics, laboratory sessions are crucial for developing students' conceptual understanding of different processes, theories and laws, problem-solving skills and lab skills. It has been emphasized that the Physics laboratories [30] are the perfect place where cooperative learning occurs. In a typical Physics lab session, students work in pairs to design and conduct experimental tasks. As a result, they have shared responsibility for all preparatory and laboratory activities, which facilitates interactive learning through discussions and mutual support. This collaborative and action-led approach typically contributes to developing deep learning processes.

**Conclusion**

This paper has discussed the benefits of Engaged pedagogy as a student-centred teaching method. In particular, it has engaged with a number of tools suitable for large group teaching in higher education, such as the Interactive lecture, Interactive lecture demonstrations, Just-in-time teaching, classroom engagement technology, Socratic Questioning and cooperative learning. It has suggested that these teaching tools can support university students' engagement with Physics more effectively than teacher-centred methods such as the traditional lecture. Additionally, the paper highlighted the importance of technology as a platform to support Active learning.